\documentclass[12pt]{article}
\let\ssection=\section
\renewcommand{\section}{\setcounter{equation}{0}\ssection}

\newcommand\mathC{\mkern1mu\raise2.2pt\hbox{$\scriptscriptstyle|$}
        {\mkern-7mu\rm C}} 
\newcommand{\mathR}{{\rm I\! R}}         

\newtheorem{definition}{Definition}[section]

\newcommand{\bra}[2][]{\ensuremath{\left._{#1}\!\left\langle{#2}\right |\right.}} 
\newcommand{\ket}[2][]{\ensuremath{\left.\left |{#2}\right\rangle\!_{#1}\right.}} 

\renewcommand{\H}{\ensuremath{{\cal H}}}
\renewcommand{\S}{\ensuremath{{\bf S}}}

\newcommand{\spec}[1]{\ensuremath{\sigma ( \hat #1)}}
\newcommand{\SP}{\ensuremath{{\bf \Sigma}}}

\newcommand\bi{\begin{itemize}}
\newcommand\ei{\end{itemize}}

\begin{document}
\begin{titlepage}
\hspace{10truecm}Imperial/TP/99-0/5

\begin{center}
{\large\bf A Topos Perspective on the
Kochen-Specker Theorem:\\[6pt] III.\ Von Neumann
Algebras as the Base Category}
\end{center}

\vspace{0.8 truecm}
\begin{center}
        J.~Hamilton\footnote{email: j.hamilton@ic.ac.uk}\\[10pt]

        C.J.~Isham\footnote{email: c.isham@ic.ac.uk}\\[10pt]

        The Blackett Laboratory\\ Imperial College of Science,
        Technology \& Medicine\\ South Kensington\\ London SW7 2BZ\\
\end{center}

\begin{center}
        J.~Butterfield\footnote{email: jb56@cus.cam.ac.uk;
            jeremy.butterfield@all-souls.oxford.ac.uk}\\[10pt] All
            Souls College\\ Oxford OX1 4AL
\end{center}
\begin{center}
        31 October, 1999
\end{center}

\vspace{0.8 truecm}

\begin{abstract}
We extend the topos-theoretic treatment given in
previous papers \cite{IB98,IB99} of assigning
values to quantities in quantum theory, and of
related issues such as the Kochen-Specker theorem.
This extension has two main parts: the use of von
Neumann algebras as a base category (Section 2);
and the relation of our generalized valuations to
(i) the assignment to quantities of intervals of
real numbers, and (ii) the idea of a subobject of
the coarse-graining presheaf (Section 3).
\end{abstract}
\end{titlepage}

\section{Introduction}
\label{Sec:Introduction} Two previous
papers---hereafter referred to as \cite{IB98,
IB99}---have developed a topos-theoretic
perspective on the assignment of values to
quantities in quantum theory. In particular, it
was shown that the Kochen-Specker theorem (which
states the impossibility of assigning to each
bounded self-adjoint operator on a Hilbert space
of dimension greater than $2$, a real number, such
that functional relations are preserved) is
equivalent to the non-existence of a global
element of a certain presheaf \SP, called the
`spectral presheaf', defined on the category $\cal
O$ of bounded self-adjoint operators on a Hilbert
space $\cal H$. In particular, the Kochen-Specker
theorem's {\em FUNC\/} condition---which states
that assigned values preserve the operators'
functional relations---turns out to be equivalent
to the `matching condition' in the definition of a
global section of the spectral presheaf. It was
similarly shown that the Kochen-Specker theorem is
equivalent to the non-existence of a global
element of a presheaf $\bf D$---called the `dual
presheaf'---defined on the category $\cal W$ of
Boolean subalgebras of the lattice ${\cal L(H)}$
of projectors on $\cal H$.

It was also shown that it {\em was\/} possible to
define so-called `generalised valuations' on all
quantities, according to which any proposition
``$A \in \Delta$'' (read as saying that the value
of $A$ lies in the Borel set of real numbers
$\Delta$) is assigned, in effect, a set of
quantities that are coarse-grainings (functions)
of $A$. To be precise, it is assigned a certain
set of morphisms in the category $\cal O$ (or
$\cal W$), the set being required to have the
structure of a {\em sieve\/}. These generalised
valuations obey a condition analogous to {\em
FUNC\/}, and other natural conditions.
Furthermore, each (pure or mixed) quantum state
defines such a valuation.

In this paper, we will extend this treatment in
two main ways. The first corresponds to our
previous concerns with the Kochen-Specker theorem
and global sections, and with generalised valuations based on sieves.
Thus, we will first discuss the issues adumbrated
above, in terms of a
base category different from $\cal O$ and $\cal
W$: namely, the category $\cal V$ of commutative
von Neumann subalgebras of an algebra of operators
(Sections \ref{Sec:Von} and \ref{Sec:sieve}).

Second, we will further develop the idea of a generalised
valuation (Section \ref{Sec:interval}). In particular, we introduce
the idea of an interval-valued valuation: at its
simplest, the idea is to assign to a quantity
$A$---not an individual member of its spectrum, as
vetoed by the Kochen-Specker theorem---but rather,
some subset of it. Though this idea seems at first
sight very different from generalised valuations,
that assign sieves to propositions ``$A \in
\Delta$'', we shall see that the two types of
valuations turn out to be closely related.

\section{Von Neumann Algebras}
\label{Sec:Von}
\subsection{Introducing $\cal V$}
We will first rehearse the definitions given in
the previous papers \cite{IB98,IB99}, of the
categories $\cal O$ and $\cal W$ defined in terms
of the operators on a Hilbert space, over which
various presheaves may be usefully constructed.
Then we will introduce a new base category $\cal
V$ which has as objects commutative von Neumann
algebras; and relate it to $\cal O$ and $\cal W$.

The categories $\cal O$ and $\cal W$ were defined
as follows. The objects of the category $\cal O$
are the bounded self-adjoint operators on the
Hilbert space $\cal H$ of some quantum system. A
morphism $f_{{\cal O}}: \hat B \rightarrow \hat A$
is defined to exist if $\hat B = f(\hat A)$ (in
the sense of \cite{IB98} Eq.\ (2.4)) for some
Borel function $f$. This category is a preorder,
and may be turned into a partially ordered set by
forming equivalence classes of operators.
Operators $\hat A$ and $\hat B$ are considered
equivalent whenever they are isomorphic in the
category $\cal O$, {\em i.e.}, when there exist
some Borel functions $f$ and $g$ such that $ \hat
B = f( \hat A)$ and $\hat A = g (\hat B)$. The
category obtained in this way is denoted $[\cal
O]$.

The category $\cal W$ is defined to have as its
objects the Boolean subalgebras of the lattice
${\cal L}({\cal H})$ of projectors on $\cal H$. A
morphism is defined to exist from $W_1$ to $W_2$
if, and only if, $W_1\subseteq W_2$. Thus $\cal W$
is just a partially ordered set (poset) equipped
with the natural categorical structure of such a
poset.

This category $\cal W$ is related to $\cal O$ via
the covariant {\em spectral algebra functor}\/
${\bf W}:{\cal O}\rightarrow\cal W$ which is
defined as follows:
\begin{itemize}
\item On objects:  ${\bf W}(\hat A):=W_A$, where
$W_A$ is the spectral algebra of the operator
$\hat A$ ({\em i.e.}, the collection of all
projectors onto the subspaces of $\cal H$
associated with Borel subsets of $\sigma(\hat
A)$).

\item On morphisms: If $f_{{\cal O}}:\hat B\rightarrow\hat A$,
then ${\bf W}(f_{{\cal O}}):W_B\rightarrow W_A$ is
defined as the subset inclusion
$i_{W_BW_A}:W_B\rightarrow W_A$.
\end{itemize}
Note that operators in the same equivalence class
in $[{\cal O}]$ will always have the same spectral
algebra.

We now wish to introduce a new base category $\cal
V$ of commutative von Neumann algebras. We first
recall (see {\em e.g.}, \cite{KR83a}) a few facts
about von Neumann algebras.

A (not necessarily commutative) von Neumann
algebra $\cal N$ is a $C$*-algebra of bounded
operators on a Hilbert space $\cal H$ which is
closed in the weak operator topology. The algebra
$\cal N$ is generated by its lattice of projectors
${\cal L}({\cal N})$, and is equal to its own
double commutant, and to the double commutant of
$\cal L(N)$:
\begin{equation}
{\cal L}({\cal N})'' = {\cal N}'' = {\cal N}
\end{equation}
The algebra contains all operators obtainable as
Borel functions $f(\hat A)$ of all normal
operators $\hat A \in \cal N$.

We now define the category $\cal V$ associated
with the Hilbert space $\cal H$ of some quantum
system. The objects $V$ in $\cal V$ are the
commutative von Neumann subalgebras of the algebra
$B({\cal H})$ of bounded operators on $\cal H$,
and the morphisms in $\cal V$ are the subset
inclusions---so if $V_2\subseteq V_1$, we have a
morphism $i_{V_2V_1}: V_2 \rightarrow V_1$. Thus
the objects in the category $\cal V$ form a poset.

The category $\cal V$ is related to $\cal O$ via a
covariant functor in a similar way to $\cal W$:
\begin{definition}\label{Defn:functorOV}
The {\em von Neumann algebra functor\/} is the
covariant functor ${\bf V}:{\cal O}\rightarrow\cal
V$ defined as follows:
\begin{itemize}
\item On objects:  ${\bf V}(\hat A):=V[A]$, where
$V[A]$ is the commutative von Neumann algebra
generated by the self-adjoint operator $\hat A$.

\item On morphisms: If $f_{{\cal O}}:\hat B\rightarrow \hat A$,
then ${\bf V}(f_{{\cal O}}):V[B]\rightarrow V[A]$
is defined as the subset inclusion
$i_{V_BV_A}:V[B]\rightarrow V[A]$.
\end{itemize}
\end{definition}

There is an even simpler relation between $\cal W$
and $\cal V$:
\begin{definition}\label{Defn:functorWV}
The {\em algebra generation functor\/} is the
covariant functor ${\bf V^W}: {\cal
W}\rightarrow\cal V$ defined as follows:
\begin{itemize}
\item On objects:  ${\bf V^W}(W):= W''$,
where $W''$ is the double commutant of $W$ in the
algebra $B({\cal H})$ of all bounded operators on
\H, so that ${\bf V^W}(W)$ is the commutative von
Neumann algebra generated by the projection
operators in $W$.

\item On morphisms: If $i_{W_2 W_1}:W_2\rightarrow W_1$ then
 ${\bf V^W}(i_{W_2 W_1}):W_2''\rightarrow W_1''$ is defined as the subset
inclusion $i_{V_2 V_1}:W_2''\rightarrow W_1''$.
\ei
\end{definition}

    The category $\cal V$ seems to give the most satisfactory
description of the ordering structure of
operators. In particular, there is no problem with
isomorphic operators: any operator isomorphic to
$\hat A$ will always be included in any subalgebra
which contains $\hat A$. Also, each von Neumann
algebra contains the spectral projectors of all
its self-adjoint members; so in a sense $\cal V$
subsumes both $\cal O$ and $\cal W$.

    Using $\cal V$ as the base category is also appealing,
from an interpretative point of view. Many
discussions and proofs of the Kochen-Specker
theorem are written in terms of sub-algebras of
operators and their relations\footnote {In
particular, Kochen and Specker in their original
paper \cite{KS67} formulate their theorem in terms
of {\em partial algebras\/} which have a similar
category-theoretic structure. Some recent work on
the `modal interpretation' focuses on certain
`beable' subalgebras of operators as those on
which valuations can be constructed
\cite{Clifton99,CH99}.}.

\subsection{Presheaves on $\cal V$}
The spectral presheaf \SP\ over $\cal O$ was
introduced in \cite{IB98}. We now define the
corresponding presheaf over $\cal V$, and the
state presheaf over $\cal V$.

\paragraph{1. The Spectral Presheaf \SP}
We recall (see {\em e.g.}, \cite{KR83a}) that the
spectrum $\sigma(V)$ of a commutative von Neumann
algebra $V$ is the set of all multiplicative
linear functionals $\kappa : V \rightarrow
\mathC$. Such a functional assigns a complex
number $\kappa(\hat A)$ to each operator $\hat A
\in V$, such that $\kappa(\hat A) \kappa (\hat B)
= \kappa (\hat A \hat B)$. If $\hat A$ is
self-adjoint, $\kappa(\hat A)$ is real and belongs
to the spectrum of the operator $\hat A$ in the
usual way.

Furthermore, $\sigma(V)$ is a compact Hausdorff
space when it is equipped with the weak-$*$
topology, which is the weakest topology such that,
for all $\hat A\in V$, the map $\tilde A
:\sigma(V)\rightarrow\mathC$ defined by
\begin{equation}
\tilde A(\kappa):= \kappa(\hat A) ,
                \label{Def:GelTransf}
\end{equation}
is continuous. The quantity $\tilde A$ defined in
Eq.\ (\ref{Def:GelTransf}) is known as the {\em
Gelfand transform\/} of $\hat A$, and the spectral
theorem for commutative von Neumann algebras
asserts that the map $\hat A\mapsto \tilde A$ is an
isomorphism of $V$ with the algebra $C(\sigma(V))$
of complex-valued, continuous functions on
$\sigma(V)$.

\begin{definition}\label{Defn:spectral}
The {\em spectral presheaf} over $\cal V$ is the
contravariant functor ${\bf \Sigma} : {\cal V}
\rightarrow {\rm Set}$ defined as follows:
\bi
\item On objects: ${\bf\Sigma}(V):=\sigma(V)$,
where $\sigma(V)$ is the spectrum of the
commutative von Neumann algebra $V$, {\em i.e.}
the set of all multiplicative linear functionals
$\kappa:V \rightarrow \mathC$.

\item On morphisms: If $i_{V_2 V_1}:V_2\rightarrow V_1$,
 so that $V_2 \subseteq V_1$, then ${\bf\Sigma}(i_{V_2
V_1}):\sigma(V_1)\rightarrow \sigma(V_2)$ is
defined by ${\bf\Sigma}(i_{V_2 V_1}) (\kappa):=
\kappa|_{V_2}$, {\em i.e.} this is the restriction
of the functionals $\kappa: V_1 \rightarrow
\mathC$ to $V_2$. \ei
\end{definition}

As discussed in \cite{IB98} for the base category
$\cal O$, the Kochen-Specker theorem may be
written in terms of the spectral presheaf. If it
existed, a global element of \SP\ over $\cal V$
would assign a multiplicative linear functional
$\kappa:V\rightarrow \mathC$ to each commutative
von Neumann algebra $V$ in ${\cal V}$ in such a
way that these functionals match up as they are
mapped down the presheaf. To be precise, the
functional $\kappa$ on $V$ would be obtained as
the restriction to $V$ of the functional $\kappa_1
: V_1 \rightarrow \mathC$ for any $V_1 \supseteq
V$.

Furthermore, when restricted to the self-adjoint
elements of $V$, a multiplicative linear
functional $\kappa$ satisfies all the conditions
of a {\em valuation}, namely:
\begin{enumerate}
\item the (real) value  $\kappa(\hat A) $ of $\hat A$ must belong to the
spectrum of $\hat A$;
\item the functional composition principle ({\em FUNC})
\begin{equation}
 \kappa(\hat B) = f(\kappa(\hat A)) \label{func}
\end{equation}
holds for any self-adjoint operators $\hat A, \hat B \in
V $ such that $\hat B = f(\hat A)$.
\end{enumerate}

The Kochen-Specker theorem, which states that no
such valuations exist on all operators on a
Hilbert space of dimension greater than two, can
therefore be expressed as the statement that the
presheaf \SP\ over $\cal V$ has no global
elements. The matching condition outlined above
therefore cannot be satisfied over the whole of
$\cal V$.

It is worth noting that a spectral presheaf may be
associated with any (non-commutative) von Neumann
algebra ${\cal N}$ by first considering the poset
${\cal A(N)}$ of all of its commutative
sub-algebras as a category, and then constructing
the functor $\SP : {\cal A} \rightarrow {\rm Set}$
in the above manner. Similar comments apply to the
other constructions introduced in the rest of this
section.

\paragraph{2. The State Presheaf ${\bf S}$}  \label{subsec:state}

A {\em state\/} $\rho $ on a $C$*-algebra ${\cal
C}$ with unit $1$ is a functional $\rho : {\cal
C}\rightarrow \mathC $ that is
\begin{enumerate}
\item{linear}
\item{positive}; {\em i.e.} $\rho(AA^*)\geq 0 $ for all $A \in {\cal C}$
\item{normalised}; so that  $\rho(1) = 1$
\end{enumerate}
(see {\em e.g.} \cite{KR83a} p. 255). The space $S$
of all states on some $C$*-algebra ${\cal C}$
of operators is a convex set whose extreme points
are the {\em pure states}. If ${\cal C}$ is
commutative, a state $\rho$ is pure if and only if
it is a multiplicative functional, {\em i.e.},
$\rho (\hat A\hat B) = \rho (\hat A) \rho (\hat
B)$ for all $\hat A,\hat B \in {\cal C}$. The set of pure
states is therefore
 the spectrum of ${\cal C}$; in particular,
the real number $\rho(\hat A)$ will belong to the
spectrum of $\hat A$ for all self-adjoint
operators $\hat A \in \cal C$.

We now define the state presheaf.
\begin{definition}\label{Defn:state}
The {\em state presheaf} \S\ over $\cal V$ is the
contravariant functor ${\bf S}: {\cal V}
\rightarrow \rm Set$ defined as follows:
\begin{itemize}
\item On objects: $\S(V)$ is the space of states  of the commutative
von Neumann algebra $V$.

\item On morphisms: If $i_{V_2 V_1}:V_2\rightarrow V_1$,
 so that $V_2 \subseteq V_1$, then ${\bf S}(i_{V_2
V_1}):\S (V_1)\rightarrow \S (V_2)$ is the
restriction of the state functionals in $\S(V_1)$
to $V_2$.
\end{itemize}
\end{definition}

A global element of this presheaf is an assignment
of a state to each commutative von Neumann
subalgebra that is consistent in the sense that
 the state on any subalgebra $V_1$ may be obtained by restriction from
 any larger subalgebra, {\em i.e.}  if $V_1 \subseteq V_2$ and $V_1
\subseteq V_3 $, then the states $\rho_2$ on $V_2$
and $\rho_3$ on $V_3$ must agree on their common
subalgebra, so that
\begin{equation}  \label{eqn:statematching}
\rho_2(\hat A) = \rho_3 (\hat A)
\end{equation}
for all $\hat A \in V_1$.

 One way to achieve this is to take a state $\rho_{\H}$ on the
(non-commutative) von Neumann algebra
$B(\H)$---for example, given by a density
matrix---and assign to each commutative subalgebra
$V$ the state $\rho_{\H}|_V$ obtained by the
restriction of $\rho_{\H}$ to $V$. However, there
may be other global elements of \S\ which are not
obtainable in this way.

The Kochen-Specker theorem tells us that the
consistency condition above cannot be satisfied
for an assignment of a {\em pure} state to each
commutative subalgebra in $\cal V$, as this would
also be a global element of \SP, and hence would
correspond to a global valuation.

\section{Sieve-Valued Generalised Valuations}\label{Sec:sieve}
In Section 2, we described the prohibition on
global assignments of real-number values to
quantum-theoretic quantities (the Kochen-Specker
theorem), in terms of the state and
spectral presheaves on $\cal V$. In this Section,
we will describe some possible {\em generalised\/}
valuations which are not excluded by the
Kochen-Specker theorem.

These constructions will have certain properties
which strongly suggest that they are appropriate
generalisations of the idea of a valuation---in
particular they satisfy a functional composition
principle analogous to Eq.\ (\ref{func}).

{\em Sieve}-valued valuations with these
properties were introduced in \cite{IB98} using
the base categories $\cal O$ and $\cal W$. These
were motivated by the observation that partial,
real-number-valued valuations ({\em i.e.}, defined
on only some subset of quantities) give rise to
such valuations, and it was shown that quantum
states could be used to define such valuations. In
\cite{IB99}, these valuations were given more
general motivations; in particular, classical
physical analogues of the quantum-theoretic
valuations were given.

This Section extends the discussions in
\cite{IB98, IB99}. First, we adapt to $\cal V$ the
definitions, results and discussion of
sieve-valued valuations. We will then look at
other possible types of generalised valuation,
namely those which are obtained by assigning to
each quantity a {\em subset\/} of its spectrum.
Some examples of these {\em interval}-valued
valuations are exhibited, and related to the
previous sieve-valued valuations.

\subsection{The Coarse-Graining Presheaf {\bf G}} \label{ssec:coarse}
We will follow \cite{IB98, IB99} in assigning
sieves primarily, not to quantities, but to
propositions about the values of quantities. In
\cite{IB98, IB99}, sieves were assigned to
propositions saying that the value of a quantity
$\hat A$ lies in a Borel set $\Delta \subset \sigma(\hat A)$, or more
precisely, to the mathematical representative of
the proposition, a projector $\hat E[A\in\Delta]$
in the spectral algebra $W_A$ of $\hat A$.

The coarse-graining presheaf was defined over
$\cal O$ (\cite{IB98}, Defn. 4.3) to show the
behaviour of these propositions as they are mapped
between the different stages of the presheaf. Thus
the coarse-graining presheaf over $\cal O$ is the
contravariant functor ${\bf G}:{\cal
O}\rightarrow{\rm Set}$ defined as follows:
\begin{itemize}
\item {\em On objects in $\cal O$:} ${\bf G}(\hat A):=
W_A$, where $W_A$ is the spectral algebra of $\hat
A$;

\item {\em On morphisms in $\cal O$:} If $f_{\cal O}:\hat B
\rightarrow\hat A$ ({\em i.e.}, $\hat B=f(\hat
A)$), then ${\bf G}(f_{\cal O}): W_A\rightarrow
W_B$ is defined as
\begin{equation}
        {\bf G}(f_{\cal O})(\hat E[A\in\Delta]):=
        \hat E[f(A)\in f(\Delta)]       \label{Def:G(O)} .
\end{equation}
\end{itemize}
Note that the action of this presheaf coarsens
propositions (and their associated projectors)
since the function $f$ will generally not be
injective and so $ \hat E[f(A)\in f(\Delta)] \ge
\hat E[A\in \Delta]$.

There are some subtleties arising from the fact
that for $\Delta$ a Borel subset of \spec{A},
$f(\Delta)$ need not be Borel. These are resolved
in \cite{IB98}, theorem (4.1), by using the fact
that if $\hat A$ has a purely discrete spectrum
(so that, in particular, $f(\Delta)$ {\em is\/}
Borel) then
\begin{equation}
\hat E[f(A)\in f(\Delta)] =\inf \{\hat Q\in
W_{f(A)}\subseteq W_A
    \mid \hat E[A\in\Delta]\leq \hat Q\}
                        \label{Theorem:inf}
\end{equation}
where the infimum of projectors is taken in the
(complete) lattice structure of all projectors on $\cal H$.
 For a
general self-adjoint operator $\hat A$ we used
this in \cite{IB98} to {\em define\/} the
coarse-graining operation; in other words, the
projection operator denoted by $\hat E[f(A)\in
f(\Delta)]$ is {\em defined\/} using the right
hand side of Eq.\ (\ref{Theorem:inf}).

This infimum construction is used again in
\cite{IB98}, Section 5.3, to define a
corresponding coarse-graining presheaf over $\cal
W$, and our construction of $\bf G$ over $\cal V$
is similar to this. Specifically, we define:
\begin{definition}\label{Defn:coarse}
The {\em coarse-graining presheaf} over $\cal V$
is the contravariant functor ${\bf G}: {\cal V}
\rightarrow {\rm Set}$ defined as follows:
\bi
\item On objects: ${\bf G}(V)$ is the lattice ${\cal L}(V)$ of projection
operators in $V$.

\item On morphisms: if $i_{V_2V_1}: V_2 \rightarrow V_1$ then ${\bf
G}(i_{V_2V_1}): {\cal L}(V_1)\rightarrow {\cal
L}(V_2)$ is the coarse-graining operation defined
on $\hat P \in {\cal L}(V_1)$ by
\begin{equation}
{\bf G}(i_{V_2V_1})(\hat P):= {\rm inf}\{\hat Q
\in {\cal L}(V_2) \mid \hat P \leq i_{V_2V_1}(\hat
Q) \}
\end{equation}
where the infimum exists because $ {\cal L}(V_2) $
is complete. \ei
\end{definition}

The coarse-graining presheaf will play a central
role in the definition of various types of
generalised valuation. To set the scene, we will
now discuss how propositions about the quantum
system behave when coarse-grained to different
stages in the base category $\cal V$. As we will
now see, this is more subtle than the analogous
process using the base category $\cal O$ or $\cal
W$ (discussed in \cite{IB98,IB99}).

In $\cal O$, a proposition about a quantum system
at some stage $\hat A \in {\cal O}$ is a statement
``$A \in \Delta$'' that the value of the quantity
$A$ lies in some Borel subset $\Delta$ of the
spectrum \spec{A} of the self-adjoint operator
$\hat A$ that represents $A$. This proposition is
associated with the stage $\hat A$ in ${\cal O}$,
and is represented by the projector $\hat E[A\in
\Delta] \in W_A$, where $W_A$ is the spectral
algebra of $\hat A$.
 For any $f_{{\cal O}}:
\hat B \rightarrow \hat A$, so that $\hat B =
f(\hat A)$, the coarse-graining operation acts on
the projector as follows:
\begin{equation}
{\bf G}(f_{{\cal O}})(\hat E[A \in \Delta]) = \hat
E[f(A) \in f(\Delta)] = \hat E[B \in f(\Delta)]
\end{equation}
where $\hat E[B\in f(\Delta)]$ is understood in
the sense explained above.

This idea of coarse-graining means that the
proposition ``$A \in \Delta$'' at stage $\hat A$
in $\cal O$ `changes' in two ways under
coarse-graining. First, the associated projector
may change---we can have $\hat E[f(A) \in
f(\Delta)] > \hat E[A \in \Delta]$. The second
change---which always occurs---is that the {\em
stage\/} of the proposition changes: after
coarse-graining, we have a proposition at stage
$f({\hat A})$ in $\cal O$, so that the
representing projector $\hat E[f(A)\in f(\Delta)]$
is thought of as belonging to $W_{f(A)}$, even if,
in fact, $\hat E[f(A) \in f(\Delta)] = \hat E[A
\in \Delta]$. In short: there is no difficulty in
interpreting a proposition (or projector) at a
stage as associated with that stage's quantity.

Similar remarks apply to $\cal W$. The main
difference from $\cal O$ is that for $\cal W$, a
stage corresponds to an equivalence class of
quantities or operators (viz.\ under the relation
of being functions of each other), not a single
quantity or operator. This just means that we
interpret a proposition ``$A \in \Delta$'', or
projector $\hat E[A \in \Delta]$, as associated
with the entire stage, {\em i.e.}, the Boolean
algebra $W_A$, not with the specific quantity $A$
or the specific operator $\hat A$. But the main
point remains the same: that with either $\cal O$
or $\cal W$, there is no difficulty in
interpreting a proposition or projector at a stage
as associated with that stage---be it a
quantity/operator; or an equivalence class, or
common spectral algebra, of such.

However, in $\cal V$, the role of propositions
such as ``$A \in \Delta$'' and the corresponding
projectors, $\hat E[A \in \Delta]$, is not so
clear. For any stage $V$ with $\hat A \in V$, the
projector $\hat E[A \in \Delta]$ will belong to
the spectral algebra of many different operators
in $V$. Indeed, it will generally also belong to
the spectral algebra of many operators {\em not\/}
in $V$, so we may write the projector as $ \hat
E[B \in \Delta_B]$ for a large number of operators
$\hat B$ with corresponding sets $\Delta_B \subset
\spec{B}$ and with $\hat B\not\in V$.

So intuition pulls in two directions. On the one
hand, similarly to $\cal W$ above, the fact that
the objects of the base category $\cal V$ are not
operators, but von Neumann algebras, prompts us to
interpret a proposition ``$A \in \Delta$'' as
associated with the entire stage, {\em i.e.}, with
the algebra $V$, rather than with the specific physical
quantity $A$ or the corresponding specific
operator $\hat A$. But on the other hand,
similarly to $\cal O$ above, operators are {\em
elements\/} of stages, so that when $\hat A$ does
belong to $V$ it seems natural to think of a
proposition ``$A \in \Delta$'' at a stage $V$ in
terms of the operator $\hat A$.

We favour the former option, since the latter
option faces difficulties when we consider
coarse-graining to a stage $V_2 \subset V_1$ which
may not contain the operator $\hat A$. To spell
these out, let us start by noting that to
understand how the proposition ``$A \in \Delta$''
at stage $V_1$ (or, more precisely, the projector
$\hat E[\hat A \in \Delta] \in {\cal L}(V_1)$)
coarse-grains to some stage $V_2 \subset V_1$,
according to Definition (\ref{Defn:coarse})
\begin{equation}
{\bf G}(i_{V_2V_1})(\hat E[A \in \Delta])= {\rm
inf}\left\{ \hat Q \in {\cal L}(V_2) \mid \hat E[A \in
\Delta] \leq i_{V_2V_1}(\hat Q) \right\},
\end{equation}
we need in general to consider three
possibilities, according to whether or not $\hat
A$ and the projector $\hat E[A \in \Delta]$ are in
$V_2$:
\begin{enumerate}
\item{$\hat A \in V_2$.} In this case, $\hat E[A \in \Delta] \in
{\cal L}(V_2) \subset V_2$, and so the projector
coarse-grains to itself: ${\bf G}(i_{V_2V_1})(\hat
E[A \in \Delta]) = \hat E[A \in \Delta]$. And
since $\hat A \in V_2$, it is natural to assign
the same interpretation---as a proposition about
the value of $\hat A$---to the coarse-grained
projector.

\item{$\hat A \notin V_2$ but  $\hat E[A \in \Delta] \in V_2$.} In
this case, the projector $\hat E[A \in \Delta]$
still coarse-grains to itself, mathematically
speaking. However, it is not clear how the
projector at stage $V_2$ could be interpreted in
terms of the value of $\hat A$, since $\hat A$ is
not present at that stage.

\item{$\hat A \notin V_2$ and  $\hat E[A \in \Delta] \notin V_2$.}
In this case, the coarse-grained projector is not
the same as the original one, and so the
proposition associated with it will no doubt be
different. It is again unclear how one could
interpret the coarse-grained projector in terms of
$\hat A$: indeed, there is in general no clear
choice as to which operator in $V_2$ is to be the
topic of the coarse-grained proposition.
\end{enumerate}

In the light of these difficulties, we will
instead adopt the first option above: we interpret
a projector $\hat P \in {\cal L}(V_1)$ as a
proposition about the entire stage $V_1$.
Formally, we can make this precise in terms of the
spectrum of the {\em algebra\/} $V_1$. That is to
say, we note that:

\begin{itemize}
\item Any projector $\hat P
 \in {\cal L}(V_1)$ corresponds not only to a subset of the spectrum
of individual operators $\hat A \in V$ (where
$\hat P \in W_A$ so $\hat P = \hat E[A \in
\Delta]$ for some $\Delta \subset \spec{A}$), but
also to a subset of the spectrum of the whole
algebra $V_1$, namely, those multiplicative linear
functionals $\kappa : V_1 \rightarrow \mathC$ such
that $\kappa (\hat P ) = 1$.

\item Coarse-graining respects this interpretation in the sense that
if we interpret $\hat P \in {\cal L}(V_1)$ as a
proposition about the spectrum of the algebra
$V_1$, then the coarse-graining of $\hat P$ to
some $V_2 \subset V_1$, given by ${\rm inf}\{ \hat
Q \in {\cal L}(V_2) \mid \hat P \leq i_{V_2V_1}(\hat Q)
\}$ is a member of ${\cal L}(V_2)$, and so can be
interpreted as a proposition about the spectrum of
the algebra $V_2$.
\end{itemize}

This treatment of propositions as concerning the
spectra of commutative von Neumann algebras,
rather than individual operators, amounts to the
semantic identification of all propositions in the
algebra corresponding to the same mathematical
projector. Thus when we speak of a proposition
``$A \in \Delta$'' at some stage $V$, with $\hat
A\in V$, we really mean the corresponding
proposition about the spectrum of the whole
algebra $V$ defined using the projector $\hat E[A
\in \Delta]$. In terms of operators, the
proposition ``$A \in \Delta$'' is {\em
augmented\/} by incorporating in it all
corresponding propositions ``$B \in \Delta_B$''
about other operators $\hat B \in V$ such that the
projector $\hat E[A \in \Delta]$ belongs to the
spectral algebra of $\hat B$, and $\hat E[A \in
\Delta] = \hat E[B \in \Delta_B]$.

\begin{definition}\label{Defn:augmented}
The {\em augmented proposition\/} at stage $V$ in
$\cal V$ associated to the projector $\hat P \in V$
is the collection of all propositions of the form
``$A \in \Delta$'' where $\hat A \in V$ and the
Borel set $\Delta\subset\mathR$ are such that
$\hat E[A \in \Delta] = \hat P$.
\end{definition}

These {\em augmented propositions} in $\cal V$
then coarse-grain in an analogous way to standard
propositions in $\cal O$: the augmented
proposition ``$A \in \Delta$'' at stage $V_1$ is
coarse-grained to ``$f(A) \in f(\Delta )$'', and
the result is an augmented proposition at the
lower stage $V_2$.

In \cite{IB98} Section 4.2.3, it was noted that
the coarse-graining presheaf over $\cal O$ was
essentially the same as the presheaf $B\SP$ over
$\cal O$, which assigns to each $\hat A$ in $\cal
O$ the Borel subsets of the spectrum of $\hat A$.
The presheaf $B\SP$ on $\cal O$ was essentially
the Borel power object of \SP, containing those
subobjects of \SP\ which are formed of Borel sets of
spectral values, with a projector $\hat E[A \in
\Delta] \in {\bf G}(\hat A)$ corresponding to the
Borel subset $\Delta \subset \SP(\hat A)$. This
connection between projectors and subsets of
spectra carries over to the algebra case.

A projection operator $\hat P \in V$ corresponds
to a subset of the spectrum of $V$, namely the set
of multiplicative linear functionals $\kappa$ on
$V$ such that $\kappa(\hat P) = 1$. Bearing in
mind that, for each $\hat A$ in $B({\cal H})$, the
function $\tilde A: \sigma(V)\mapsto \mathC$
given by the Gelfand transform of $\hat A$,  $\tilde A(\kappa) :=\kappa(\hat A)$, is
continuous, we see that the subset of the spectrum
of $\sigma(V)$ that corresponds to $\hat P$ is
{\em closed\/} in the compact Hausdorff topology
of $\sigma(V)$.

In fact, we can say more than this since, by
virtue of the spectral theorem for commutative
$C^*$-algebras, the operator $\hat P\in V$ is
represented by a function $\tilde
P:\sigma(V)\rightarrow\mathR$. Since $\hat
P^2=\hat P$, we see that $\tilde P$ is necessarily
the characteristic function, $\chi_P$ say, of some
subset of $V$, namely the set of all multiplicative
linear functionals $\kappa$ on $V$ such that
$\kappa(\hat P)=1$. However, by virtue of the
spectral theorem, $\chi_P$ is in fact a {\em
continuous\/} function from $\sigma(V)$ to
$\{0,1\}\subset\mathR$, and hence the subset
concerned is both open and closed. Thus the subset
of $\sigma(V)$ corresponding to a projection
operator $\hat P\in V$ is a {\em clopen\/} subset
in the spectral topology. Conversely, of course,
each clopen subset of $\sigma(V)$ corresponds to a
projection operator $\hat P$ whose representative
function $\tilde P$ on $\sigma(V)$ is the
characteristic function of the subset.

So in analogy with $B\SP$ on $\cal O$, we may
define a similar presheaf on $\cal V$, which we
will denote ${\rm Clo}{\bf\Sigma}$:
\begin{itemize}
\item On objects: ${\rm Clo}\SP(V)$ is defined to be the
set of clopen subsets of the spectrum $\sigma(V)$ of the
algebra $V$; each such clopen set is the set of
multiplicative linear functionals $\kappa$ such
that $\kappa (\hat P) = 1$ for some projector
$\hat P \in V$.

\item On morphisms: for $V_2 \subset V_1$, we define
\begin{equation}
{\rm Clo}\SP(i_{V_2V_1})\left( \{ \kappa \in
\sigma(V_1) \mid \kappa (\hat P)= 1\} \right) = \{
\chi \in \sigma(V_2) \mid \chi ({\bf
G}(i_{V_2V_1})(\hat P))= 1\}
\end{equation}
\end{itemize}
There is an isomorphism between $\bf G$ and ${\rm
Clo}\SP$, and so we can think of $\bf G$ on
$\cal V$ as being the `clopen' power object of
\SP\ on $\cal V$.

\subsection{Sieve-Valued Generalised Valuations} \label{ssec:sievegenval}
In view of the discussion of propositions in the
previous subsection, our definition of a
sieve-valued generalised valuation for the base
category $\cal V$ will define the valuations on
projectors in the explicit context of an algebra
$V$. That is to say, the truth-value associated
with a projector $\hat P$ depends on the context
of a particular algebra $V$ containing $\hat P$.
As in any topos of presheaves (cf.\ \cite{IB98},
Appendix), the subobject classifier ${\bf \Omega}$
in the topos $\rm Set^{{\cal V}^{{\rm op}}}$ is a
presheaf of sieves. Since ${\cal V}$ is a poset,
sieves may be identified with lower sets in the
poset. We define ${\bf \Omega}$ as follows:
\begin{itemize}
\item {On objects:
${\bf\Omega}(V)$ is the set of sieves in $\cal V$
on $V$;

We recall that ${\bf\Omega}(V)$ has (i) a minimal
element, the empty sieve, $0_V = \emptyset$, and
(ii) a maximal element, the principal sieve,
true$_V = \downarrow\!_V:=\{V' \mid V' \subseteq V\}$.
}

\item On morphisms: ${\bf \Omega}(i_{V_2V_1}):{\bf \Omega}(V_1)
\rightarrow {\bf \Omega}(V_2)$ is the pull-back of
the sieves in ${\bf \Omega}(V_1)$ along
$i_{V_2V_1}$ defined by:
\begin{eqnarray}
{\bf \Omega}(i_{V_2V_1})(S) = i_{V_2V_1}^* (S)
&:=& \{ i_{V_3V_2}:V_3 \rightarrow V_2 \mid
i_{V_2V_1} \circ i_{V_3V_2} \in S \}\\
    &=&\{V_3\subset V_2|V_3\in S\}
\end{eqnarray}
\noindent for all sieves $S \in {\bf \Omega}(V_1)$.
\end{itemize}

Then we define:
\begin{definition}\label{Defn:gen-val-V}
A {\em sieve-valued generalised valuation\/} on
the category $\cal V$ in a quantum theory is a
collection of maps $\nu_V:{\cal L}(V) \rightarrow
{\bf\Omega}(V)$, one for each `stage of truth' $V$
in the category $\cal V$, with the following
properties:
\end{definition}
\noindent {\em (i) Functional composition}:
\begin{eqnarray}
\lefteqn{\mbox{\ For any ${\hat P} \in {\cal
L(V)}$ and any $V' \subseteq V$ so that }
i_{V'V}:V' \rightarrow V} \hspace{3cm}\nonumber
\\[3pt]
        &&\nu_{V'}({\bf G}(i_{V'V}(\hat P))= i_{V'V}^*(\nu_V(\hat P)).
\hspace{2cm} \ \label{FC-gen-V}
\end{eqnarray}

\noindent {\em (ii) Null proposition condition}:
\begin{equation}
                \nu_V(\hat 0)=0_V \label{Null-gen-V}
\end{equation}

\noindent {\em (iii) Monotonicity}:
\begin{equation}
        \mbox{If }\hat P,\hat Q\in {\cal L}(V)\mbox{ with }
        \hat P\leq\hat Q,\mbox{ then }
        \nu_V(\hat P)\leq\nu_V(\hat Q). \label{Mono-gen-V}
\end{equation}

\noindent We may wish to supplement this list
with:

y\noindent\smallskip
{\em (iv) Exclusivity}:
\begin{equation}
        \mbox{If $\hat P,\hat Q\in {\cal L}(V)$ with
$\hat P\hat Q=\hat 0$ and $\nu_V(\hat P)= {\rm
true}_V$, then $\nu_V(\hat Q)< {\rm true}_V$}
\label{Excl-gen-V}
\end{equation}
and
\smallskip\noindent
{\em (v) Unit proposition condition}:
\begin{equation}
        \nu_V(\hat 1)=\mbox{true}_V.
        \label{unit-prop-cond-V}
\end{equation}

Note that in writing Eq.\ (\ref{FC-gen-V}), we
have employed Definition \ref{Defn:coarse} to
specify the coarse-graining operation in terms of
an infimum of projectors, as motivated by Theorem
4.1 of \cite{IB98}.

The topos interpretation of these generalised
valuations remains as discussed in Section 4.2 of
\cite{IB98} and Section 4 of \cite{IB99}. Adapting
the results and discussion to the category $\cal
V$, we have in particular the result that because
of the $FUNC$ condition, Eq.\ (\ref{FC-gen-V}),
the maps $N_V^{\nu}:{\cal L}(V) \rightarrow {\bf
\Omega}(V)$ defined at each stage $V$ by:
\begin{equation}
    N_V^{\nu}(\hat P) = \nu_V(\hat P)
\end{equation}
define a natural transformation $N^{\nu}$ from
$\bf G$ to ${\bf \Omega}$. Since ${\bf \Omega}$ is
the subobject-classifier of the topos of
presheaves, ${\rm Set}^{\cal V^{\rm op}}$, these
natural transformations are in one-to-one
correspondence with subobjects of $\bf G$; so that
each generalised valuation defines a subobject of
$\bf G$. We will pursue this topic in more detail
in Section \ref{ssec:semsubs}.

\subsection{Sieve-Valued Valuations Associated with Quantum States}
 \label{ssec:genvalquantumstate}
We recall (for example, \cite{IB98}, Definition
4.5) that each quantum state $\rho$ defines a
sieve-valued generalised valuation on $\cal O$ or
$\cal W$, in a natural way. For example, on $\cal
O$, the generalised valuation was defined as
\begin{eqnarray}
 \nu^\rho(A\in\Delta)&:=&\{f_{\cal O}:\hat B\rightarrow \hat A
\mid {\rm Prob}(B\in f(\Delta);\rho)=1\}\nonumber
\\[2pt]
                &\,=&\{f_{\cal O}:\hat B\rightarrow \hat A
\mid {\rm tr}(\rho\,\hat E[B\in f(\Delta)])=1\}
\label{Def:nurhoDelta}
\end{eqnarray}
 Thus the generalised valuation associates
to the proposition all arrows in $\cal O$ along
which the projector corresponding to the
proposition coarse-grains to a projector which is
`true' in the usual sense of having a Born-rule
probability equal to $1$. This construction is
easily seen to be a sieve, and satisfies
conditions analogous to Eqs.\
(\ref{FC-gen-V}--\ref{unit-prop-cond-V}) for a
generalised valuation on ${\cal O}$ (\cite{IB98}, Section
4.4).

We also recall that there is a one-parameter
family of extensions of these valuations, defined
by relaxing the condition that the proposition
coarse-grains along arrows in the sieve to a
`totally true' projector. That is to say, we can define the
sieve
\begin{eqnarray}
 \nu^{\rho,r}(A\in\Delta)&:=&\{f_{\cal O}:\hat B\rightarrow \hat A
\mid {\rm Prob}(B\in f(\Delta);\rho) \ge r
\}\nonumber
\\[2pt]
                &\,=&\{f_{\cal O}:\hat B\rightarrow \hat A
\mid {\rm tr}(\rho\,\hat E[B\in f(\Delta)]) \ge
r\} \label{Def:nurho-r-Delta}
\end{eqnarray}
 where the proposition ``$A\in\Delta$'' is only required to
coarse-grain to a projector that is true with
some probability greater than $r$, where $0.5 \leq
r \leq 1$.

Furthermore, if one drops the exclusivity
condition, one can allow probabilities less than
0.5, i.e. $0 < r < 0.5$.

We now introduce the same kind of valuation using
$\cal V$ as the base category.

As was discussed in Section \ref{ssec:coarse} in
relation to $\bf G$, when using the base category
$\cal V$ it is more natural to interpret a
projector $\hat P \in {\cal L}(V)$ as a
proposition about the spectrum of the commutative
subalgebra $V$, rather than about the value of
just one operator. Such an {\em augmented
proposition} at a stage $V$ in $\cal V$ will
correspond to a projector $\hat P \in {\cal L}(V)$, and can
be thought of as the family of propositions
``$A \in \Delta$'' for all $\hat A \in V$
that have $\hat P $ as a member of their spectral
algebra with $\hat E[ A \in \Delta] = \hat P$.

So we define a sieve-valued generalised valuation
associated with a quantum state $\rho$ as follows:

\begin{definition}\label{Defn:gen-val-V-rho}
The sieve-valued valuation $ \nu^{\rho}_{V_1}$ of
a projector $\hat P \in V_1$ associated with a
quantum state $\rho$ is defined by:
\begin{equation}
\nu^{\rho}_{V_1} (\hat P) := \{ i_{V_2V_1}: V_2
\rightarrow V_1 \mid \rho \: [{\bf
G}(i_{V_2V_1})(\hat P)] = 1 \} \label{eqn:nurhoV}
\end{equation}
\end{definition}
This assigns as the truth-value at stage $V_1$ of
a projector $\hat P \in V_1$, a sieve on $V_1$
containing (morphisms to $V_1$ from) all stages
$V_2$ at which $\hat P$ is coarse-grained to a
projector which is `totally true' in the usual
sense of having Born-rule probability $1$.

One readily verifies that Eq.\ (\ref{eqn:nurhoV})
defines a generalised valuation in the sense of
Definition \ref{Defn:gen-val-V}. The verification
is the same, {\em mutatis mutandis}, as for
generalised valuations on $\cal O$, given in
Section 4.4 of \cite{IB98}. As an example, we take
the functional composition condition. Again, this
requires that the sieves pull back in the
appropriate manner; if $V_2\subseteq V_1 $ and
hence $i_{V_2V_1}: V_2 \rightarrow V_1$, then:
\begin{eqnarray}
i_{V_2V_1}^* (\nu^{\rho}_{V_1} (\hat P)) &:= & \{
i_{V_3V_2}: V_3 \rightarrow V_2 \mid i_{V_2V_1}
\circ i_{V_3V_2} \in \nu^{\rho}_{V_1}(\hat P)
 \} \\ & = & \{ i_{V_3V_2}: V_3 \rightarrow V_2 \mid \rho
[{\bf G}(i_{V_2V_1} \circ i_{V_3V_2} )(\hat P)] =
1 \}
\end{eqnarray}
whereas
\begin{equation}
\nu^{\rho}_{V_2} ({\bf G}(i_{V_2V_1})(\hat P)) :=
\{ i_{V_3V_2}: V_3 \rightarrow V_2 \mid \rho [{\bf
G}(i_{V_3V_2})(\, {\bf G}(i_{V_2V_1})(\hat P)\, )
] = 1\}
\end{equation}
and hence {\em FUNC} is satisfied since ${\bf
G}(i_{V_2V_1} \circ i_{V_3V_2} )(\hat P) = {\bf
G}(i_{V_3V_2})({\bf G}(i_{V_2V_1})(\hat P))$.

Again, we can obtain a one-parameter family of
such valuations by introducing a probability $r$:
\begin{equation}
\nu^{\rho ,r}_{V_1} (\hat P) := \{ i_{V_2V_1}: V_2
\rightarrow V_1 \mid \rho \: [{\bf
G}(i_{V_2V_1})(\hat P)] \ge r \}
\label{eqn:nurhoV-r}
\end{equation}

\section{Interval-Valued Generalised Valuations} \label{Sec:interval}
\subsection{Introducing Interval-Valued Valuations}
The sieve-valued generalised valuations on $\cal
V$ discussed in Section \ref{Sec:sieve}, and their
analogues on $\cal O$ and $\cal W$ (discussed in
\cite{IB98, IB99}) are one way of assigning a
generalised truth value to propositions in a way
that is not prevented by the Kochen-Specker
theorem. We are now going to investigate another
possibility---of assigning sets of real numbers to
operators---and relate it to our generalised
valuations. An algebra $V$ in $\cal V$ will be
assigned a subset of its spectrum, {\em i.e.} a
set of multiplicative linear functionals on $V$
which corresponds to a subset of the spectrum of
each operator in the algebra.

We will call such assignments `interval
valuations', as the primary motivation for these assignments is
the wish to assign some interval of real numbers
to each operator. Note that, in the latter
context, `interval' means just some subset of
$\mathR$, that is to say, an interval in our sense
need not be a connected subset.

Despite the marked difference between
interval-valued and sieve-valued valuations
---projectors or propositions {\em versus\/} algebras as
arguments, and sieves {\em versus\/} sets of real
functionals as values---it turns out that these
two kinds of valuations are closely related.

\subsection{Subobjects of \SP} \label{ssec:subobjSP}
We now consider assigning to each algebra $V$ in
$\cal V$ a clopen subset $I(V) \subseteq
\sigma(V)$. This set $I(V)$ of multiplicative
linear functionals $\kappa$ on $V$ leads to an
assignment to each self-adjoint operator $\hat
A\in V$ of a subset of the spectrum of the
operator, $\Delta_A
 =  \{ \kappa(\hat A) \mid \kappa \in I(V) \}
 \subseteq \spec{A}$,
in such a way that the appropriate relationships
between these subsets are obeyed, so that if $\hat
B = f (\hat A)$, we have $\Delta_B = f(\Delta_A)
$.
For we note that $\Delta_A$ is equal to $\tilde
A[I(V)]$, where the Gelfand transform $\tilde
A:\sigma(V)\rightarrow\mathR$ is defined in Eq.\
(\ref{Def:GelTransf}). However, $I(V)$ is a closed
subset of the compact Hausdorff space $\sigma(V)$,
and hence is itself compact. Then, since $\tilde
A:\sigma(V)\rightarrow\mathR$ is continuous, it
follows that $\tilde A[I(V)]$ is also compact;
thus $\Delta_A$ is a {\em compact\/} subset of
$\mathR$.

On the face of it, the fact that $\Delta_A$ is
compact might appear problematical for functions
$f:\sigma(\hat A)\rightarrow\mathR$ that are Borel
but not continuous (since the image of a compact
set by a Borel function need not itself be
compact). In effect, the problem is that $f\circ
\tilde A:\sigma(V)\rightarrow\mathR$ may not be
continuous, even though it is supposed to
represent the operator $\hat B=f(\hat A)$.
However, a more careful study of the spectral
theorem for a commutative von Neumann algebra
shows that the Borel function $f\circ\tilde A$ can
be replaced by a unique continuous function ({\em
i.e.}, $\tilde B$) without changing anything of
significance in the algebraic structure
\footnote{For example, see the discussion on p.324
in \cite{KR83a}.}. In addition, there exists some
continuous function $\tilde
f:\mathR\rightarrow\mathR$ such that $\tilde
B=\tilde f\circ \tilde A$, and the natural-looking
equation $\Delta_B=f(\Delta_A)$ is to be
understood as $\Delta_B=\tilde f(\Delta_A)$ where
necessary.

One of course expects to require such an
interval-valued assignment, $I$, to obey some
version of $FUNC$ along morphisms in the category
$\cal V$. The most obvious version is, for $V_2
\subset V_1$:
\begin{equation}
I(V_2) = I(V_1)|_{V_2} \label{Defn:FUNCeq}
\end{equation}
 so that the functionals in $I(V_2)$ are just the restriction
to $V_2$ of those in $I(V_1)$.

But from the perspective of the theory of
presheaves, it is natural to take such assignments
$I$ as given by subobjects of \SP. These clearly
exist---for example there is the trivial subobject
assigning to each algebra the whole of its
spectrum. One can think of the
Kochen-Specker theorem as restricting the
`smallness' of the subobjects: we cannot take a
subobject that consists of a singleton set at each
stage, since that would be a global element. If
${\bf I}$ is a subobject of \SP, it will obey (by the
definition of `subobject') a {\em weaker\/}
version of $FUNC$: viz.,
\begin{equation}
{\bf  I}(i_{V_2V_1})({\bf I}(V_1)) := {\bf I}(V_1)|_{V_2} \subseteq {\bf  I}(V_2)
 \label{Defn:FUNCsset}
\end{equation}

\paragraph{Example 1: The `true' subobject of \SP\ arising from a quantum state}
One subset of the spectrum of an operator $\hat A$
which arises naturally from a quantum state $\rho$ is the
set of values which can occur in a measurement of
$\hat A$ when the system is in the state $\rho$. This set $\Delta_A$ is
given in the following way.

For any pair of projectors $\hat E[A\in\Delta],
\hat E[A\in\Delta'] \in W_A$, such that ${\rm
tr}(\rho\hat E[A \in \Delta]) = {\rm tr}(\rho\hat
E[A\in\Delta']) = 1$, there is a smaller projector
$\hat E[A\in\Delta'']=\hat E[A\in\Delta]\hat
E[A\in\Delta']$ (where
$\Delta''=\Delta\cap\Delta'$) such that ${\rm tr}(
\rho\hat E[A\in\Delta'']) = 1$, and so we may form
a descending net of projectors whose infimum
exists, and belongs to $W_A$, since $W_A$ is
complete. Thus there is a smallest projector $\hat
E_{\rm min}$ which is of the form $\hat
E[A\in\Delta_A]$ for some subset $\Delta_A$ of the
spectrum of $\hat A$. Furthermore, the net of
projectors converges strongly (and therefore
weakly) to $\hat E[A\in\Delta_A]$, and therefore,
since---for fixed $\rho$---the map $B({\cal
H})\rightarrow\mathC$ given by $\hat A\rightarrow
{\rm tr}(\rho\hat A)$, is weakly continuous, it
follows that ${\rm tr}(\rho \hat E[A\in\Delta_A])
= 1$. This operator $\hat E_{\rm min}=\hat
E[A\in\Delta_A]$ is sometimes called the {\em
support\/} of $\hat A$ in the state $\rho$.

For an algebra $V$, we can define the
corresponding subset of its spectrum by taking
those multiplicative linear functionals on $V$
which assign the value $1$ to each `true
projector', {\em i.e.}, those projectors in
$T^\rho(V) := \{\hat P \in V \mid {\rm tr}(\rho
\hat P) = 1\} \subset {\cal L}(V)$. The connection
to the above assignment to operators is clear; for
if $\hat A \in V$, then $\Delta_A = \{ \kappa(\hat
A) \mid \kappa(\hat P) = 1, \; \; \forall \hat P
\in T^\rho(V)\}$.

So the corresponding `true subobject' of \SP\ is
given by:
\begin{equation}
{\bf I}^{\rho}(V) = \{ \kappa \in \sigma(V) \mid \kappa
(\hat P) = 1, \; \; \forall \hat P \in T^\rho(V)
\}. \label{eqn:truesubobj}
\end{equation}
Again, since the map $\hat P \rightarrow {\rm tr}(\rho \hat P)$ is
weakly continuous, the infimum $\hat Q := {\rm inf}\{ \hat P \in
T^\rho(V)\}$ is in $T^\rho(V)$. Since $\kappa(\hat
Q) = 1 $ implies $\kappa(\hat P) = 1$ for all
$\hat P \in T^\rho(V)$, the above construction may
be written as the clopen set
\begin{equation}
{\bf I}^{\rho}(V) := \{\kappa \in \sigma(V) \mid \kappa
(\hat Q) = 1\} \label{eqn:truesubobjQ}
\end{equation}
and so for non-trivial $\hat Q$ this will
necessarily be a proper subset of $\sigma(V)$
(cf.\ the discussion at the end of Section
\ref{ssec:coarse} on the connection between
projectors and subsets of spectra, in relation to
${\rm Clo}{\bf\Sigma}$).

The above construction describes a subobject of
\SP\ since if we have $V_2 \subset V$ then
\begin{equation}
{\bf I}^{\rho}(V_2) = \{\kappa \in \sigma(V_2) \mid
\kappa (\hat P) = 1 \; \; \forall \hat P \in
T^\rho(V_2) \}
\end{equation}
and since $T^\rho(V_2) = \{ \hat P \in V_2 \mid
{\rm tr}(\rho \hat P) = 1\} \subseteq T^\rho(V)$,
we have that $\hat Q_2 \ge \hat Q$. Therefore $\{
\kappa \mid \kappa(\hat Q) = 1 \} \subseteq \{
\kappa \mid \kappa(\hat Q_2) = 1 \}$, and hence
${\bf I}^\rho(V)|_{V_2}\subseteq {\bf I}^{\rho}(V_2) $ as
required for the subobject condition
(\ref{Defn:FUNCsset}) to hold.

\paragraph{Example 2: Subobjects of \SP\ from sieve-valued valuations}
Given an extra condition on a sieve-valued
valuation $\nu$, the above construction of the
`true subobject' of \SP\ can be adapted to use
$\nu$ to define an interval-valuation $I^{\nu}$
via:
\begin{equation}
{\bf I}^{\nu}(V) = \{ \kappa \in \sigma(V) \mid \kappa
(\hat P) = 1, \; \; \forall \hat P \in T^\nu(V) \}
\label{eqn:Inu}
\end{equation}
where now we define
\begin{equation}
T^\nu(V) := \{ \hat P \in {\cal L}(V) \mid
\nu_{V}(\hat P) = {\rm true}_{V} \}.
\label{eqn:TnuSP}
\end{equation}

We now look at the conditions under which this
gives a subobject of \SP. We note first that Eq.\
(\ref{eqn:Inu}) is not in general a proper subset
of $\sigma(V)$. It will be empty unless the
infimum $\hat Q$ of the set of projectors
$T^\nu(V)$ is non-zero; for we can again write
${\bf I}^{\nu}(V)$ in the form of the clopen set
\begin{equation}
{\bf I}^{\nu}(V) = \{ \kappa\in\sigma(V) \mid
\kappa(\hat Q) = 1 \}
\end{equation}
which must be empty if $\hat Q = \hat 0$, the zero
projector.

The condition on $\nu$ for this construction to be
a subobject is a matching condition on this
infimum:
\begin{equation} \mbox{ For } V_1 \subset V \mbox{  we
require that }\hat Q_1 \ge \hat Q
\label{eqn:infcond}
\end{equation}
where $\hat Q$ and $\hat Q_1$ are the infima of
the sets $T^\nu(V)$ and $T^\nu(V_1)$ respectively,
defined as above. For, if this is not the case,
there will be some $\kappa\in\sigma(V)$ such that
$\kappa (\hat Q) = 1$ (and hence $\kappa\in
{\bf I}^{\nu}(V)$) but $\kappa(\hat Q_1) < 1$. Therefore
$\kappa|_{V_1} \notin {\bf I}^{\rho}(V_1)$, and the
subobject condition Eq.\ (\ref{Defn:FUNCsset})
will not be satisfied.

A sufficient condition for Eq.\
(\ref{eqn:infcond}) to hold is that if
  $ V_1 \subset V$ then
$T^\nu(V_1) \subseteq T^\nu(V)$. This condition is
satisfied for the valuation $\nu^\rho$ arising
from a quantum state in the probability one case
via Eq.\ (\ref{Def:nurhoDelta}), and the resulting
construction is, of course, the same as the
Example 1 above.

The condition also holds for valuations from a
quantum state $\nu^{\rho,r}$ where a probability
$r$ is introduced, as in Eq.\
(\ref{eqn:nurhoV-r}). In that case, however, there
will be many algebras $V\in {\cal V}$ where there
is no non-trivial infimum to the set $T^\nu(V)$,
and so the corresponding subobject of \SP\ will be
the empty set over much of $\cal V$.

\subsection{Global Elements of \bf G} \label{ssec:globG}
As was discussed at the end of Section
\ref{ssec:coarse}, there is a natural
interpretation of the coarse-graining presheaf
$\bf G$ as a subobject of the power object of \SP,
{\em i.e.}, a presheaf of subobjects of \SP.

Given this interpretation of $\bf G$, it is
natural to take interval-valued valuations as
given by global elements of ${\bf G}$. A global
element $\gamma$ of $\bf G$ obeys, for $V_2
\subset V_1$:
\begin{equation}
\gamma(V_2) ={\bf G}(i_{V_2V_1})(\gamma(V_1)),
\end{equation}
which is a $FUNC$ condition with an equality like
Eq.\ (\ref{Defn:FUNCeq}) (as opposed to the subset
version Eq.\ (\ref{Defn:FUNCsset}) required for
subobjects of \SP).

Recalling that for any projector selected by
$\gamma$ at stage $V_1$, {\em i.e.} $\gamma(V_1) =
\hat P \in {\cal L}(V_1)$, the action of $\bf G$
is ${\bf G}(i_{V_2V_1}): \hat P \mapsto {\rm
inf}\left\{ \hat Q \in {\cal L}(V_2) \mid \hat P
\leq i_{V_2V_1}(\hat Q) \right\}$, we see that any
such global element $\gamma$ defines an assignment
$I^{\gamma}$ to each algebra of a subset of its
spectrum by:
\begin{equation}
I^{\gamma}(V_1) := \{ \kappa \in \sigma(V_1) \mid
\kappa (\gamma(V_1))= 1 \}
\end{equation}
and, since $\hat P\leq {\bf G}(i_{V_2V_1})(\hat
P)$, we see that $\kappa(\hat P)=1$ implies
$\kappa({\bf G}(i_{V_2V_1})(\hat P))=1$, so that
$I^\gamma(V_1)|_{V_2}\subseteq I^\gamma(V_2)$, and
hence this construction is a subobject of \SP\ as
well as an interval-valued valuation satisfying
the stronger version of $FUNC$, Eq.\
(\ref{Defn:FUNCeq}).

In this way, an interval-valued
valuation---assigning a subset of the spectrum at
each stage---is given by every global element of
{\bf G}.

As in Example 2 of the previous Subsection,
certain sieve-valued valuations can be used to
give global elements of ${\bf G}$. The natural way
to define such a global element is by:
\begin{eqnarray}
\gamma^\nu(V_1) &:=& {\rm inf} \{ \hat P \in {\cal
L}(V_1) \mid
  \nu_{V_1}(\hat P) = {\rm true}_{V_1} \}  \\
     &\; =:&\hat Q_1 = {\rm inf} \{ \hat P \mid \hat P \in
T^\nu(V_1) \} \label{eqn:gammanu}
\end{eqnarray}
where again, as in Eq.\ (\ref{eqn:TnuSP}), $T^\nu(V_1) := \{ \hat P
\in {\cal L}(V_1) \mid \nu_{V_1}(\hat P) = {\rm true}_{V_1}
\}$.

The matching condition for this to be a global
element of ${\bf G}$ is stronger than for it to be
a subobject of \SP: we now require that for every
$V_2 \subset V_1$,
\begin{equation}
{\rm inf} \{ \hat P \in T^\nu(V_2) \} =: \hat Q_2 =
{\bf G}(i_{V_2V_1}) \left( {\rm inf} \{\hat P \in
T^\nu(V_1) \} \right) = {\bf G}(i_{V_2V_1})(\hat
Q_1).
\end{equation}

The case of a valuation arising from a quantum
state for the probability one case---Example 1 in
the previous Subsection---may be cast in this form.
For that case we have that if $V_2 \subset
V_1$ then $T^\rho(V_2) = \{ \hat P \in V_2 \mid
{\rm tr}(\rho\hat P) = 1\} \subseteq T^\rho(V_1)$, and so each $\hat
P_2 \in V_2 $ is given by ${\bf G}(i_{V_2V_1})(\hat P_1)$ for
some $\hat P_1 \in V_1$, since in particular this is true when  $\hat
P_1,\hat P_2$ denote the same projector thought of as belonging to $V_1$
and $V_2$ respectively. Then since $\hat P_1 \le \hat P_1'$ implies
${\bf G}(i_{V_2V_1})(\hat P_1) \le {\bf G}(i_{V_2V_1})(\hat P_1')$, it
follows that
${\rm inf}\{\hat P\in T^\rho(V_2)\} = {\bf
G}(i_{V_2V_1}) ({\rm inf}\{ \hat P\in
T^\rho(V_1)\})$, and so we have a global element of
$\bf G$.

The valuations $\nu^{\rho,r}$ arising from
a quantum state via the probability $r$
construction (Eq. (\ref{eqn:nurhoV-r})), while
giving subobjects of \SP, do not satisfy the
stronger condition necessary to form global
elements of ${\bf G}$.

One way to avoid these issues arising from having to take infima of certain
sets of projectors is to look at subobjects of
$\bf G$, rather than its global elements.

\subsection{Subobjects of {\bf G}}  \label{ssec:subG}
As described at the end of Section
\ref{ssec:sievegenval}, every sieve-valued
valuation $\nu$ on projectors induces a
morphism $N^\nu: \bf G \rightarrow\Omega$ in the
topos of presheaves over $\cal V$, and hence
corresponds to a subobject of $\bf G$. This
subobject ${\bf T}^\nu$ is given at each stage $V$
by the set
\begin{equation}
{\bf T}^\nu (V) = \{ \hat P \in {\cal L}(V) \mid
\nu_V(\hat P) = {\rm true}_V \}
  \label{eqn:Tnu}
\end{equation}

This is the same construction as
Eq.(\ref{eqn:TnuSP}), but now considered as
defining a subobject of $\bf G$. In this subsection
we discuss such subobjects, in particular those
arising from sieve-valued valuations associated
with a quantum state $\rho$ for the probability 1
and probability $r$ cases, as defined in equations
(\ref{eqn:nurhoV}) and (\ref{eqn:nurhoV-r}).

\subsubsection{The probability 1 case}
To obtain a subobject of ${\bf G}$ from a quantum
state $\rho$, we take at each stage $V_1$ the
subset of the lattice of projectors ${\cal
L}(V_1)$
\begin{equation}
{\bf T}^{\rho}(V_1) := \{\hat P \in V_1 \mid {\rm
tr}(\rho\hat P) = 1 \} \; \; \subset {\cal
L}(V_1).
\end{equation}
This is, of course, the subobject arising via Eq.\
(\ref{eqn:Tnu})
 from the sieve-valued valuation $\nu^\rho$ associated
with the quantum state $\rho$, defined in Eq.\
(\ref{eqn:nurhoV}).

This forms a subobject of the coarse-graining
presheaf since if $i_{V_2V_1}:V_2 \rightarrow V_1$
then for any $\hat P \in {\cal L}(V_1)$
\begin{equation} \label{eqn:rhoG}
{\rm tr}(\rho ({\bf G}(i_{V_2V_1})(\hat P))) \ge
{\rm tr}(\rho\hat P).
\end{equation}
Hence it follows that
\begin{equation}
{\bf G}(i_{V_2V_1})({\bf T}^{\rho}(V_1)) \subseteq
{\bf T}^{\rho}(V_2)
\end{equation}
and therefore ${\bf T}^{\rho}$ is a genuine
subobject of ${\bf G}$. Also $ {\cal L}(V_2)
\subseteq {\cal L}(V_1)$ and for all $\hat P_2 \in
{\cal L}(V_2)$, ${\bf G}(i_{V_2V_1})(\hat P_1) =
\hat P_2 $ where $\hat P_1 , \hat P_2$ denote the
same projector $\hat P$ thought of as belonging to
${\cal L}(V_1)$ and ${\cal L}(V_2)$ respectively.
It follows that for all $\hat P_2 \in {\bf
T}^{\rho}(V_2)$ we have $\hat P_1\in {\bf
T}^{\rho}(V_1) $, and therefore
\begin{equation}
{\bf T}^{\rho}(V_2) \subseteq {\bf
G}(i_{V_2V_1})({\bf T}^{\rho}(V_1))
\end{equation}
and so this subobject obeys a functional
composition principle $ {\bf T}^{\rho}(V_2) = {\bf
G}(i_{V_2V_1})({\bf T}^{\rho}(V_1))$ with an equality.

In this particular case, we can use the subobject
${\bf T}$ of ${\bf G}$ to give at each stage a subset
$I^\rho(V)$ of the spectrum of the von Neumann
algebra $V$ via
\begin{equation}
I^{\rho}(V) := \{ \kappa \in \SP(V) \mid
\kappa(\hat P)= 1 \: \: \forall \hat P \in {\bf T}^{\rho}(V)
\}\subset \SP(V)
\end{equation}

Since the lattice ${\cal L}(V)$ is complete, there
exists a smallest projector in the set
${\bf T}^{\rho}(V)$ as in the case of the global element of $\bf G$:
 $\hat Q := {\rm inf} \{\hat P \mid \hat P \in {\bf T}^{\rho}(V)\}$.
This is then the single projector at each stage
defined by Eq.\ (\ref{eqn:gammanu}) for the case
where $\nu$ is the sieve-valuation associated with
the state $\rho$, and so the global element
construction may be recovered in this case.

This then corresponds to a single augmented
proposition at each stage, namely the collection
of propositions ``$A \in \Delta $'' for each
operator $ \hat A \in V$ and Borel subset $\Delta$ such that $\hat E[A \in
\Delta] = \hat Q$.

\subsubsection{The Probability $r$ Case}
As discussed in Section \ref{ssec:globG}, not
all sieve-valuations lead to global elements of
$\bf G$, or to subobjects of \SP. In particular,
there is no global element of $\bf G$ associated with the
probability $r$ case for a quantum state $\rho$.
However, we can take a subobject of $\bf G$:
\begin{equation}
{\bf T}^{\rho,r}(V) := \{\hat P \in V \mid {\rm
tr}(\rho \hat P) \ge r \}
\end{equation}
where $0 \le r \le 1$. Heuristically, this is the
set of propositions in ${\cal L}(V)$ which are
assigned a probability greater than $r$. Equation
(\ref{eqn:rhoG}) still holds, and hence this still
gives a genuine subobject of ${\bf G}$.

The propositions $\{\hat P \in V \mid {\rm
tr}(\rho\hat P)\ge r\}$ do not in general have a
non-trivial infimum, (except when $r=1$), and the
infima will not obey the necessary matching
conditions to allow us to construct a global
element of ${\bf G}$ in the same way as was
possible in the $r=1$ case.

\subsubsection{Semantic Subobjects} \label{ssec:semsubs}
This discussion of certain subobjects of $\bf G$
induced via Eq.\ (\ref{eqn:Tnu}) by generalised
valuations $\nu$ prompts us to ask what
characterises subobjects of $\bf G$ which can be
induced in this way. The defining conditions for a
generalised valuation  $\nu$, Eqs.
(\ref{FC-gen-V}--\ref{Excl-gen-V}), give four
properties of the corresponding subobject ${\bf
T}^\nu$ of {\bf G}, defined according to Eq.\
(\ref{eqn:Tnu}).

\begin{enumerate}
\item {\em Functional composition:}
this ensures that we have a genuine subobject
${\bf T}^\nu$ of $\bf G$, since it implies in
particular that for all $\hat P \in {\bf T}^\nu(V_1)$,
so that $\nu(\hat P) = {\rm true}_{V_1}$, it is
true that $\nu_{V_2} \left( {\bf
G}(i_{V_2V_1})(\hat P) \right) = {\rm
true}_{V_2}$, so $ {\bf G}(i_{V_2V_1})({\bf T}^\nu(V_1))
\subseteq {\bf T}^\nu(V_2)$.

\item{\em Null proposition condition:}
this implies that the always-false projector $\hat
0_V$ never belongs to ${\bf T}^\nu(V)$.

\item{\em Monotonicity:}
for $\hat P,\hat Q \in {\cal L}(V)$ with $\hat
P\leq\hat Q$, this condition implies that if $\hat
P \in {\bf T}^\nu(V)$ then also $\hat Q \in {\bf T}^\nu(V)$;
so ${\bf T}^\nu(V)$ is required to be an upper set in
${\cal L}(V)$.

\item{\em Exclusivity}:
if $\hat P,\hat Q\in {\cal L}(V)$ with $\hat P\hat
Q=\hat 0$ and $\hat P \in {\bf T}^\nu(V)$ then
 $\hat Q \notin {\bf T}^\nu(V)$.
\end{enumerate}
Note that 1.\ and 2.\ are consequences of Eq.\
(\ref{FC-gen-V}) and (\ref{Null-gen-V})
respectively, and weaker than them. On the other
hand, 3.\ and 4.\ are equivalent to Eq.\
(\ref{Mono-gen-V}) and Eq.\ (\ref{Excl-gen-V})
respectively.

We may wish to relax the exclusivity condition
here (and the corresponding condition for
generalised valuations) depending on the type of
valuation being studied. As has already been
pointed out, the exclusivity condition is not
satisfied (and would not be expected to hold) for
valuations where the probability required for a
proposition to belong to the corresponding
subobject of $\bf G$ is less than a half.

Given the properties listed in 1.--4., we
can then turn this around, and define a {\em
semantic subobject\/} of $\bf G$ as being a
subobject satisfying these properties. Semantic
subobjects then form a set of possible generalised
valuations for our quantum theory.

\subsection{Interval-Valued Valuations from Ideals}
We now present another way of obtaining an
interval-valued valuation on $\cal V$ which does
not rely on the use of sieve-valuations and the
coarse-graining presheaf.

We recall that the closed ideals in a commutative
von Neumann algebra $V$ are in one-to-one
correspondence with the closed subsets of the
spectrum $\sigma(V)$ of $V$. More precisely,
according to the spectral theorem, the algebra $V$
is isomorphic to the algebra $C(\sigma(V))$ of
continuous, complex-valued functions on its
spectrum; and a closed ideal in the algebra
corresponds to the set of all functions in
$C(\sigma(V))$ that vanish on the associated
closed subset of $\sigma(V)$.

Furthermore, according to a general result about
extremely disconnected Hausdorff spaces (which
$\sigma(V)$ is), each closed subset of $\sigma(V)$
differs from a unique clopen set by a meagre set.
Therefore, any assignment of a closed ideal to
each algebra in $\cal V$ will produce an
interval-valued valuation, assigning to each $V$
in $\cal V$ a clopen subset $\Xi$ of its spectrum.
The functions in the ideal are then precisely
those that are disjoint to the characteristic
function of $\Xi$, {\em i.e.}, those operators in
$V$ orthogonal to the projector $\hat P$ that
corresponds to the characteristic function of
$\Xi$.

Furthermore, there is a natural way to make such assignments of closed
ideals to each algebra $V$ in $\cal V$. We note
that an ideal $\iota$ in any non-commutative von
Neumann algebra $\cal N$ will induce an ideal
$\iota_V \subset V $ in each of its commutative
subalgebras $V$ by restriction: $\iota_V = \iota
\cap V$. In this way, the ideal $\iota$ assigns to each
$V$ the clopen subset of $\sigma(V)$ associated
with $\iota_V$.

We note that the lattice of closed, two-sided
ideals in any non-commutative algebra $\cal N$ is an example of a quantale
(\cite{MP99}). This structure can be viewed as the
analogue of a spectrum for the non-commutative von
Neumann algebra $\cal N$, and as we have just shown, it
represents another natural collection of
interval-valued valuations in our framework.

\paragraph{Example:}

The subset $\iota^{\psi}$ of the set $B(\cal H)$
of bounded operators on some Hilbert space $\cal
H$ defined by
\begin{equation}
\iota^{\psi} = \{\hat A \in B({\cal H}) \mid \hat
A \psi = 0 \}
\end{equation}
forms a closed left ideal since $\hat B \hat A
\psi = 0$ for all $\hat B \in B(\cal H)$, $\hat A
\in \iota^{\psi}$. This induces a two-sided ideal
in each commutative von Neumann subalgebra $V$:
\begin{equation}
\iota^{\psi}(V) = \{\hat A \in V \mid \hat A \psi
= 0 \}.
\end{equation}

We will now show that this ideal corresponds to
the `true' subobject of \SP\ described in Example
1 of Section \ref{ssec:subobjSP} for the quantum state
$\psi$.

We will denote the set of projectors in
$\iota^\psi(V)$ by ${\cal P}(\iota^\psi(V))$. If
the projectors $\hat Q_1,\hat Q_2 \in {\cal
P}(\iota^\psi(V))$, so that $\hat Q_1 \psi = 0 =
\hat Q_2 \psi$, then we also have $(\hat Q_1 +
\hat Q_2) \psi = 0$, so $(\hat Q_1 + \hat Q_2) \in
{\cal P}(\iota^\psi(V))$. It follows that there
exists a largest projector $\hat Q_{\rm max} \in
\iota^{\psi}(V)$
\begin{equation}
\hat Q_{\rm max} = {\rm sup}{\cal
P}(\iota^\psi(V)).
\end{equation}

As was noted above, a projector in $V$ corresponds
to a characteristic function in $C(\sigma(V))$ of
some clopen subset $\Xi\subset\sigma(V)$, and
the set of functions disjoint to a characteristic
function are the functions in the ideal. The ideal
in $V$ is then associated with the subset of
$\sigma(V)$ for which the projector $(\hat 1 -
\hat Q_{\rm max})$ corresponds to the
characteristic function {\em i.e.}, the set
$\{\kappa \in \sigma(V) \mid \kappa (\hat 1 - \hat Q_{\rm max})
= 1\}$. We also note that $(\hat 1 - \hat Q_{\rm
max})\psi = \psi$, and so $\bra{\psi}(\hat 1 -
\hat Q_{\rm max})\ket{\psi} = 1$, and indeed
$(\hat 1 - \hat Q_{\rm max})$ is the smallest
projector with this property. This therefore
corresponds to the subobject of \SP\ given by
${\bf I}^\rho$ in Eq.\ (\ref{eqn:truesubobjQ}) for the
state $\rho$ on $V$ given by $\rho(\hat A) =
\bra{\psi} \hat A \ket{\psi}$.

\section{Conclusions}

In this paper, we have extended the
topos-theoretic perspective on the assignment of
values to quantities in quantum theory to the base
category $\cal V$ of commutative von Neumann
algebras; a category which generalizes the
categories of contexts $\cal O$ and $\cal W$ used
in \cite{IB98,IB99}. As we have seen, the main
results of \cite{IB98,IB99}---both the leading
ideas, and the mathematical constructions---can be
adapted to the von Neumann algebra case. (Though
we have not spelt out all the details of this
adaptation, piece by piece, one can check the
details for the topics we have omitted---for
example, classical analogues---and general
motivations for sieve-valued valuations, as
discussed in \cite{IB99}.)

This adaptation of our results to $\cal V$ is
straightforward, except that, as noted in Section
3.1, we need to be careful about two issues: (i)
about interpreting a proposition ``$A \in \Delta$"
relative to a von Neumann algebra $V$ as a
context; and (ii) about the properties of spectral
topologies. The upshot is that:
(i) we interpret a projector $\hat P$ at a context
$V$ in terms of all ``$A \in \Delta$", with $A \in
V$ and ${\hat E}[A \in \Delta] = {\hat P}$; and
(ii) $\bf G$ is isomorphic to the clopen power
object ${\rm Clo}{\bf\Sigma}$ of the spectral
presheaf ${\bf \Sigma}$.

Accordingly, we conclude that the base category
$\cal V$ of commutative von Neumann algebras is as
natural a basis for developing the topos-theoretic
treatment of the values of quantities in quantum
theory, as were our previous categories $\cal O$
and $\cal W$; indeed, in certain respects $\cal V$
is a better basis, since it includes the others in
a natural way. In particular, it is a natural
basis for addressing the various topics listed in
Section 6 of \cite{IB98}.

 In this paper, we have addressed one such topic, using $\cal V$: that
of interval-valued valuations. As we saw in
Section 4, there are close
connections between our sieve-valued valuations,
and the assignment to quantities of subsets of
their (operators') spectra, and so also the assigments to commutative
algebras of subsets of their spectra. We displayed these
connections in various ways: in terms of
subobjects of $\bf \Sigma$, in terms of global
elements of $\bf G$, and in terms of subobjects of
$\bf G$. One main idea in making these connections
was the set $T^{\nu}(V)$ of projectors ${\hat P}
\in V$ that are ``wholly true'' according to the
sieve-valued valuation $\nu$: which for the case
of $\nu$ given by a quantum state $\rho$ is
essentially the familiar quantum-theoretic idea of
the support of a state. Finally, we noted in
Section 4.5 that once we use $\cal V$, rather than
$\cal O$ or $\cal W$, as our base category of
contexts, we can use ideals in the von Neumann
algebras to define interval-valued valuations in a
natural way: though again, care is needed about
the spectral topologies.

\end{document}